\documentclass{wqcd03}                 

\usepackage{amssymb}
\usepackage{amsmath}

\newcommand{\tr}{{\mathrm{tr}}}
\newcommand{\TR}{{\mathrm{Tr}}}

\confname{QCD@Work 2003 - International Workshop on QCD, Conversano, Italy, 
14--18 June 2003}

\title{Variational analysis of deconfinement in gluodynamics\thanks{In memory of  Ian Kogan: friend, mentor and colleague.}}

\author{J.~Guilherme~Milhano\addressmark{a}\addressmark{b} }

\address[a]{Department of Physics and Astronomy, Vrije Universiteit, The Netherlands}

\address[b]{CENTRA, Instituto Superior T\'ecnico, Universidade T\'ecnica de Lisboa, Portugal}

\begin{document}

\begin{abstract}
The deconfinement transition in 3+1 dimensional gluodynamics is studied using the gauge invariant variational method introduced by Kogan and Kovner a few years ago. We identify a first order phase transition, characterized by a discontinuous jump in the entropy of the system, resulting in a transparent picture of the mechanism of deconfinement. The calculation of the ratio of the transition temperature to the mass of the lightest glueball in the model yields 0.18 in complete agreement with the lattice estimate. 
\end{abstract}

\maketitle


\section{Introduction}

In the almost 25 years since the pioneering work of Polyakov \cite{polyakov} and Susskind \cite{susskind} much effort has been devoted to attempts to understand both the basic physics and quantitave features of the deconfining phase transition of QCD.

The high temperature phase, i.e. for temperatures well above the critical temperature $T_c$, is becoming well understood, and is widely believed to resemble a plasma of almost free quarks and gluons.
However, the transition region, $T_c<T<2 T_c$, is very poorly understood. This region is the most interesting one since it is there that the transition between `hadronic' and `partonic' degrees of freedom occurs.

The study of the transition region is a complicated and inherently non-perturbative problem which has mostly evaded treatment by analytical methods.
The variational method introduced several years by Kogan and Kovner \cite{kk} appears well suited to a discussion of such a problem.

Recently, the method was applied to the study of the phase transition in  $SU(N)$ gluodynamics.
As a result of these studies \cite{Kogan:2002yr,bg}, we identified a first order phase transition at a transition temperature, in units of the lighest state in the theory, of $T_c=0.18$,  in complete agreement with the lattice estimate for $SU(3)$ \cite{Teper:1998kw}. 

We minimize the relevant thermodynamic potential at finite temperature, i.e. the Helmholtz free energy, on a set of suitably chosen trial density matrix functionals.
As with any variational calculation there is an element of guesswork involved. However, the choice of the trial states is dictated by rather general principles. As in the quantum mechanical case, the trial state ought to be simultaneously rich enough to span the interesting physics and to lead to calculable results. This last requirement is particularly stringent, since only functional integrals of the gaussian form  can be explicitly evaluated. Specific to the quantum field case are the requirements of gauge invariance and of correct UV behaviour. To understand this last statement, it suffices to mention first, that the free energy is dominated by UV modes while we are interested in the IR physics and, second, that in an asymptotically free theory such as  gluodynamics, low and high momentum modes are, in general, non-separable.
The method of \cite{kk,Kogan:2002yr,bg} to be discussed below accommodates all these requirements, leading to what is, to the author's best knowledge, the only first principle analytical analysis of the deconfinement phase transition in gluodynamics.

\section{The variational Ansatz}

We consider density matrices which, in the field basis, have Gaussian matrix elements and where gauge invariance is explicitly imposed by projection onto the gauge-invariant sector of the Hilbert space
\begin{align} \label{qcdans}
\rho &[A,A^{'}] =  \nonumber\\
&=\int DU \; \exp\bigg\{-\frac{1}{2}\int_{x,y}
\Big[ A_{i}^{a} (x) G^{-1 ab}_{ij}(x,y) A_{j}^{b}(y) \nonumber\\
&\hspace{2.5cm}+ A^{'Ua}_{i} (x) G^{-1 ab}_{ij}(x,y) A^{'Ub}_{j}(y)\nonumber\\
&\hspace{2.5cm}- 2 A_{i}^{a} (x) H^{ab}_{ij}(x,y) A^{'Ub}_{j} (y) \Big] \bigg\}\, ,
\end{align}
where $\int_{x,y}=\int d^3 x\, d^3 y$, $DU$ is the $SU(N)$ group-invariant measure, and under an $SU(N)$ gauge transformation $U$
\begin{gather} \label{gt}
A^{a}_{i} (x) \rightarrow A^{U a}_{i} (x) = S^{ab} (x) A^{b}_{i} (x) + \lambda^{a}_{i} (x)\, ,
\end{gather}
with 
\begin{align}
S^{ab} &= \frac{1}{2} \mathrm{tr} ( \tau^a U^{\dag} \tau^b U )\, ,\\ 
\lambda^{a}_{i} &= 
\frac{i}{g} \mathrm{tr} ( \tau^a U^{\dag} \partial_i  U )\, ,
\end{align} 
and 
$\frac{\tau^a}{2}$ form an $N \times N$ Hermitian representation of $SU(N)$:
$[ \frac{\tau^a}{2} , \frac{\tau^b}{2} ] = i f^{abc} \frac{\tau^c}{2}$ with normalization 
$\mathrm{tr} ( \tau^a \tau^b ) = 2 \delta^{ab}$.

We take the variational functions diagonal in both colour and Lorentz indices, and translationally invariant
\begin{align}
    G^{-1ab}_{ij} (x,y) &= \delta^{ab} \delta_{ij}
    G^{-1}(x-y)\, ,\nonumber\\
H^{ab}_{ij} (x,y) &= \delta^{ab} \delta_{ij}
    H(x-y)\, .
\end{align}
Further, we split the momenta into high and low modes with $k \lessgtr M$ and restrict
the kernels $G^{-1}$ and $H$ to the one parameter momentum space forms
\begin{align} \label{rkers}
G^{-1}(k) &= \left\{ \begin{matrix} M, \; k<M \\k, \; k>M \end{matrix}\right. \, , \nonumber \\
H (k) &= \left\{ \begin{matrix} H, \; k<M \\0, \; k>M \end{matrix}\right.\, .
\end{align}

The logic behind this choice of ansatz is the following. At finite
temperature we expect $H(k)$ to be roughly proportional to the
Bolzmann factor $\exp\{-{\mathsf E}(k)\beta\}$. In our ansatz, the role of
one particle energy is played by the variational function
$G^{-1}(k)$ and its form is motivated by the propagator of a massive scalar field, i.e. $(k^2+M^2)^{1/2}$. We will be  interested only in temperatures close to
the phase transition, and those we anticipate to be small,
$\mathsf{T_{c}}\le M$. For those temperatures one particle modes
with momenta $k\ge M$ are not populated, and we thus put $H(k)=0$.
For $k\le M$ the Bolzmann factor is non-vanishing, but small.
Further, it depends only very weakly on the value of the momentum.
With the above restrictions on the kernels,  only two variational parameters, $M$ and $H$, remain.

Importantly, the density matrix functional in eq.~(\ref{qcdans}) describes, for $H=0$, a pure state $\rho=|\Psi [A]><\Psi [A]|$ where $\Psi [A]$ are Gaussian wave functionals. For $H\neq 0$, eq.~(\ref{qcdans}) describes a mixed state with $|H|$ proportional to the entropy of the trial density matrix.

The expectation value of a gauge invariant operator in the variational state eq.~(\ref{qcdans}) is then given by
\begin{align}
\label{eq:expvalue}
    \langle {\cal O}\rangle_{A,U}
    & = Z^{-1}\TR (\rho {\cal O}) \nonumber  \\
    = &Z^{-1} \int   {\cal D}U{\cal D}A
    \:{\cal O}(A,A')\cdot\nonumber\\
	&\quad\cdot \exp \bigg\{ -\frac{1}{2} \Big[ A G^{-1} A +  {A'}^U G^{-1}{A'}^U\nonumber\\
         &\hspace{3cm}- 2A H {A'}^U \Big]\bigg\}\Bigg|_{A'=A}\, ,
\end{align}
where $Z$ is the normalization of the trial density matrix $\rho$, i.e.
\begin{align}
    Z&=\TR\rho \nonumber\\
	&= \int {\cal D}U{\cal D}A \cdot\nonumber\\
	&\cdot\exp \bigg\{
    -\frac{1}{2} \Big[
    A G^{-1} A +  {A}^U G^{-1}{A}^U
    -2 A H {A}^U \Big]\bigg\} 
\end{align}

To evaluate the above expressions we first perform, for fixed $U(x)$, the gaussian integration over the vector potential $A$. For $Z$ we get, in leading order in $H$,
\begin{multline}
    Z=  \int {\cal D}U \exp \bigg\{
    -\frac{1}{2}\lambda\Big(
    \frac{G^{-1}}{2} + \frac{H}{4} (S+S^T)
    \Big)\lambda\\
+\frac{3}{4} HG\:\tr (S+S^T)\bigg\}\, .
\end{multline}

We now integrate out the high momentum, $k^2>M^2$,  modes of $U$ perturbatively to one-loop order.
This effects a renormalization group transformation on the low modes, replacing the bare coupling $g^2$ by the running coupling $g^2(M)$ \cite{kk,brko}.
To one-loop accuracy, the coupling $g^2(M)$ runs identically to the Yang-Mills coupling \cite{brko}.

\section{The effective $\sigma$-model}

The normalization $Z$ can be then interpreted as the generating functional 
\begin{equation}
    Z=\TR\rho =  \int {\cal D}U e^{- {\cal S}(U)}
\end{equation}
for an effective non-linear $\sigma$-model for the low momentum modes ($k^2<M^2$) defined by the action
\begin{align}
    \label{eq:action}
    {\cal S}(U) &=\frac{M}{2 g^2} \tr( \partial U\partial U^\dagger)\nonumber\\
   &\quad - \frac{H}{8 g^2}\tr\Big[(U^\dagger \partial U - \partial U^\dagger U)
    (\partial U U^\dagger - U\partial U^\dagger) \Big]\nonumber\\
    &\quad- \frac{1}{4\pi^2} H M^2 \tr U^\dagger \tr U\, ,
\end{align}
where $U$ independent pieces have been dropped.

The matrix $U$ plays the same role as Polyakov's loop $P$ at
finite temperature --- the functional integration over $U$ projects
out the physical subspace of the large Hilbert space on which the
Hamiltonian of gluodynamics is defined.

This $\sigma$-model has a  phase transition \cite{kk} at the critical point (for $\Lambda_{QCD} = 150$Mev, $N=3$ and  with the one-loop Yang-Mills $\beta$ function)
\begin{equation}
M_c=\Lambda_{QCD}e^{\frac{24}{11}}=8.86\Lambda_{QCD}=1.33 \mathrm{Gev}\, .
\label{mc}
\end{equation}
For $M<M_c$, the $\sigma$-model is in a disordered, $SU(N)_L\otimes SU(N)_R$ symmetric, phase with massive excitations and where $\langle U\rangle = 0$. Since $U$ is  the Polyakov loop, this corresponds to a confined state. When $M>M_c$, the $\sigma$-model is in an ordered, $SU(N)_V$ symmetric, phase with massless Goldstone bosons for which $\langle U\rangle \neq 0$, corresponding to a deconfined state.
With this analysis we have established a correspondence between the $\sigma$-model phase transition and the deconfinement transition in $SU(N)$ gluodynamics. 

In fact, this correspondence can be argued in rather general terms.
Instead of restricting ab initio the density matrix to the form eq.~(\ref{qcdans}), 
imagine that we take some arbitrary gauge-invariant density matrix ansatz that depends on the $A$ fields and is integrated over the $U$ fields.
We allow this new ansatz (and whatever kernels it may contain) to remain arbitrary until we have no choice but to restrict it.
Then we integrate out the A fields to obtain a partition function of $U$ fields with respect to some action.

Next we introduce a separation of
momenta into high and low modes with $k \lessgtr M$
and integrate out the high mode $U$ fields as before. This effects a renormalisation group transformation on the low modes,
replacing the bare coupling $g^2$ --- which is not arbitrary, since it is defined by the gauge transformations eq.~(\ref{gt}) --- by the running coupling
$g^2(M)$. 
Now provided our ansatz is sufficiently close to the correct density matrix for $SU(N)$, the theory will be asymptotically free.
We are thus left with an action for the low
modes which is, once again,  a complicated $\sigma$-model, with a renormalised coupling $g^2(M)$ which we expect to be small provided $M$
is large, and vice versa.

Now consider this model as a statistical mechanical model at `temperature' $g^2(M)$. We make the plausible assumption 
that this $\sigma$-model will, as $M$ is varied, undergo a symmetry-breaking transition
at `temperature' $g^2(M_c)$ from a `thermally disordered' (symmetric) phase at large $g^2(M_c)$ to an ordered phase at small $g^2(M_c)$.
Further, it is clear --- since the Polyakov loop $\langle U \rangle$ is zero in the former phase and non-zero in the latter ---
that this $\sigma$-model phase transition corresponds directly to the deconfinement transition in the $SU(N)$ theory. 

On review, it is clear that our only assumptions are that the ansatz is sufficiently close to $SU(N)$
and that the low mode $\sigma$-model undergoes a symmetry-breaking phase transition. In particular, let the ansatz, which is arbitrary and need not be Gaussian, be the \emph{correct} density matrix for $SU(N)$. The first assumption is certainly true. 
If the second assumption is also true, then we have constructed an exact argument that the deconfinement transition in $SU(N)$
corresponds to the phase transition in the low mode $\sigma$-model.

Thus, in order to study deconfinement in $SU(N)$, our aim should be to model the physics of each $\sigma$-model phase as 
accurately as possible and calculate the transition scale $M_c$. We then
calculate the free energy of $SU(N)$ in each phase, 
including any possible contribution from the high modes, at temperature $T$ and extract the minimal free energy. The deconfinement 
transition occurs
at the temperature for which the free energies calculated in the ordered and disordered phases of the low mode $\sigma$-model coincide.

Although we will take eq.~(\ref{qcdans}) as the ansatz for the density matrix, we shall keep the kernels $G^{-1}$ and $H$ arbitrary until we have no choice but to restrict them.

\section{Calculation of the free energy}

The Helmholtz free energy ${\sf F}$ of the density matrix
$\rho$ is given by
\begin{equation}
    {\mathsf F} = \langle {\mathsf H} \rangle
     - {\mathsf T}   \langle{\mathsf S} \rangle\, ,
    \end{equation}
where ${\sf H}$ is the standard Yang-Mills Hamiltonian
\begin{equation}
{\mathsf H}= \int d^{3}x \left[{1\over 2}E^{a2}_i+{1\over
2}B^{a2}_i\right]\, , \label{ham}
\end{equation}
with
\begin{align}
E^a_i(x)&=i{\delta\over \delta A^a_i(x)}\, , \nonumber \\
B^a_i(x)&={1\over 2}\epsilon_{ijk}
\{\partial_jA_k^a(x)-\partial_kA^a_j(x)\nonumber\\
&\hspace{2cm}+gf^{abc}A_j^b(x)A_k^c(x)\}\, ,
\end{align}
${\sf S}$ is the entropy, and ${\sf T}$ is the
temperature.

Thus
\begin{equation}
\label{eq:freeenergy}
     {\mathsf F}
    = \frac{1}{2}\Big(
    \TR ({E}^2\varrho)
    +\TR ({B}^2\varrho)
    \Big)
    + {\mathsf T}\cdot\TR( \varrho \ln\varrho) \, .
\end{equation}

In the disordered phase, no progress seems possible without restricting the arbitrary kernels.  Following \cite{Kogan:2002yr},
we adopt the forms eq.~(\ref{rkers}).
For small $H$, we consider only the first non-trivial order in $H$, that is a term of $o(H\ln H)$ in the entropy. This term can be written as a product of  left $SU(N)$ and  right $SU(N)$ currents and does, therefore, vanish in the disordered, $SU(N)_L\otimes SU(N)_R$ symmetric, phase \cite{Kogan:2002yr}. The remaining contribution to the free energy, the average of the Hamiltonian, is evaluated in the mean field approximation \cite{kk}. The free energy is minimized for $M=M_c\simeq 1.33 \mathrm{GeV}$
\begin{gather} \label{minlf}
F_{dis} = - \frac{N^2 M_{c}^{4}}{30 \pi^2}\, .
\end{gather}

The simplest option to evaluate the free energy in the disordered phase is to use perturbation theory. Perturbation is certainly appropriate for large enough values of $M$, where the expectation value of the $U$ field is close to unity. From numerical studies \cite{kogut} it is known that the transition occurs when the expectation value of $U$ is greater than $.5$. We can thus expect perturbation theory to be qualitatively reliable down to the transition point. 
In the leading order perturbation theory approximation to the ordered phase of the  $\sigma$-model, however, minimisation with respect to arbitrary kernels $G^{-1}$ and $H$ for both high and low modes is possible. Further, the analysis can be carried out to all orders in the thermal disorder kernel $H$.

In this approximation, the $U$ matrices can be parameterised in the standard exponential form and expanded in the coupling $g$
\begin{gather}
U = \exp\bigg\{ig \varphi^a \frac{\tau^a}{2}\bigg\} = 1 + ig \varphi^a \frac{\tau^a}{2} + \dots  
\end{gather}
Hence at leading order one can take
\begin{align}
U &\simeq 1\, , \nonumber \\
\partial_i U &\simeq  ig \partial_i \varphi^a \frac{\tau^a}{2}. 
\end{align}
Thus, the gauge transformations (\ref{gt}) reduce to
\begin{gather}
A^{a}_{i} \rightarrow A^{a}_{i} - \partial_i \varphi^a  
\end{gather}
and the Hamiltonian (\ref{ham}) reduces to 
\begin{gather}
\mathcal{H} = \frac{1}{2} \left[ E^{a2}_{i} + (\epsilon_{ijk} \partial_j A^{a}_{k})^2 \right]\, . 
\end{gather}
These last two equations describe the theory $U(1)^{N^2-1}$: in the leading order of $\sigma$-model perturbation theory, the 
$SU(N)$ Yang--Mills theory reduces to the $U(1)^{N^2-1}$ free theory.
Moreover, the density matrix eq.~(\ref{qcdans}) becomes Gaussian again, because the gauge transformations are linear. One has
\begin{multline} \label{gans}
\rho [A,A^{'}] = \int D\varphi \; 
\exp\bigg\{-\frac{1}{2} \Big[ A G^{-1} A \\
+ (A' - \partial \varphi) G^{-1} (A' - \partial \varphi) \\
- 2 A H (A' - \partial \varphi) \Big] \bigg\}\, . 
\end{multline}

Now the theory of $N^2-1$ $U(1)$ free fields in $3+1$ dimensions 
is completely tractable; the variational analysis for the $U(1)$ theory
(with Gaussian ansatz (\ref{gans})) was discussed in \cite{Gripaios:2002xb}. The free energy in momentum space in terms of the arbitrary kernels
$G^{-1}$ and $H$ is
\begin{align}
F =& \frac{N^2-1}{2} \int \frac{d^3p}{(2\pi)^3} 
\bigg[  G^{-1}(1 + GH) + p^2 G (1 - GH)^{-1} \nonumber\\
 &\quad- 4T \Big(  \log \Big[ \frac{GH}{  (1-(GH)^2)^{1/2} - (1-GH)}\Big] \nonumber\\
&\qquad- \log \Big[ \frac{1- (1-(GH)^2)^{1/2}}{GH}\Big] \cdot\nonumber\\
&\qquad\quad\cdot\Big( \frac{1- (1-(GH)^2)^{1/2}}{(1-(GH)^2)^{1/2}-(1-GH)} \Big) \Big) \bigg]\, .
\end{align}  
The kernels which minimise the free energy are
\begin{align} \label{mkers}
G^{-1} &= p 
\left( \frac{1+ e^{-\frac{2p}{T}}}{1 - e^{-\frac{2p}{T}}}\right), \nonumber \\
H &= 2p
 \left( \frac{e^{-\frac{p}{T}}}{1 - e^{-\frac{2p}{T}}}\right)
\end{align}
and the minimal value of the free energy at temperature $T$ is
\begin{gather}
F = - \frac{\pi^2 N^2 T^4}{45}.
\end{gather}

So we see that the free energy of $SU(N)$ is minimised with $M=M_c$ in the disordered phase of the $\sigma$-model for temperatures from zero up to
a temperature $T_c$ where
\begin{gather} 
F = - \frac{N^2 M_{c}^{4}}{30 \pi^2} =  - \frac{\pi^2 N^2 T_{c}^{4}}{45}\, ,
\end{gather}
which in turn implies 
\begin{gather}
T_c = \left( \frac{3}{2}\right)^{1/4} \frac{M_c}{\pi} \simeq 470 \mathrm{MeV}\, .
\end{gather}

\section{Conclusions}

We find that the deconfinement phase transition is strongly first order with a transition temperature of  $470 \mathrm{MeV}$.
The transition is due to an abrupt jump in the entropy. In the  disordered phase it is zero since glueballs are heavy and, therefore, their contribution is suppressed by the Boltzmann factor $\exp(-M_g/T)$. In the ordered phase, the entropy is non-vanishing and proportional to the number of coloured states.
 
The major uncertainty in the calculation is due to the fact that the scale $M_c$ is very sensitive to the mean field approximation. It is, therefore, only meaningful to compare the results of our approach with  lattice estimates for $M_c$ independent  quantities. For one such quantity, the transition temperature in units of the mass of the lightest glueball in the model, i.e. $2M_c$ \cite{Gripaios:2002bu}, we find
\begin{gather} \label{ratio}
\frac{T_c}{2M_c} = \frac{1}{2\pi}(\frac{3}{2})^{1/4} \simeq 0.18.
\end{gather}
which is in agreement with the lattice result.

\section*{Acknowledgments}
This talk is based upon work done in collaboration with Ben Gripaios,  Ian Kogan and Alex Kovner. This work and my participation in the workshop were funded by the European Commission IHP programme under contract HPRN-CT-2000-00130.

\end{document}